\documentclass[11pt,a4paper]{article}
\usepackage{amsmath,amsthm,amssymb,amscd}
\usepackage{latexsym}
\usepackage{indentfirst}
\usepackage{bibentry}
\usepackage{textcomp}
\usepackage{float}
\usepackage{cases}
\usepackage{multirow}
\usepackage{booktabs}
\usepackage{graphicx}
\usepackage{color}
\usepackage[numbers,sort&compress]{natbib}
\usepackage{booktabs}
\usepackage{authblk}
\usepackage[numbers]{natbib}
\setcitestyle{open={},close={}}
\makeatletter \@addtoreset{equation}{section}

\makeatother
\newtheorem{thm}{Theorem}[section]

\theoremstyle{definition}

\begin{document}

\title{\Large Painlev\'{e} V for a Jacobi unitary ensemble with random singularities}

\author[1,2]{{\small Mengkun Zhu}\footnote{zmk@qlu.edu.cn}}
\author[3,4]{{\small Chuanzhong Li}\footnote{corresponding author: lichuanzhong@nbu.edu.cn}}
\author[2]{{\small Yang Chen}\footnote{corresponding author: yayangchen@um.edu.mo}}
\affil[1]{\small School of Mathematics and Statistics, Qilu University of Technology (Shandong Academy of Sciences)\\
Jinan 250353, China}
\affil[2]{\small Department of Mathematics, University of Macau,
Avenida da Universidade, Taipa, Macau, China}
\affil[3]{\small College of Mathematics and Systems Science, Shandong University of Science and Technology, Qingdao 266590, China}
\affil[4]{\small School of Mathematics and Statistics, Ningbo University,
Ningbo 315211, China}

\renewcommand\Authands{ and }

\date{}
\maketitle

\begin{abstract}
In this paper, we focus on the relationship between the fifth Painlev\'{e} equation and a Jacobi weight perturbed with random singularities,
\begin{equation*}
w(z)=\left(1-z^2\right)^{\alpha}{\rm e}^{-\frac{t}{z^2-k^2}},~~~z,k\in[-1,1],~\alpha,t>0.
\end{equation*}
By using the ladder operator approach, we obtain that an auxiliary quantity $R_n(t)$, which is closely related to the recurrence coefficients of monic polynomials orthogonal with $w(z)$, satisfies a particular Painlev\'{e} V equation.
\end{abstract}

{\small Keywords: Jacobi unitary ensembles, Orthogonal polynomials, Painlev\'{e} V, Ladder operators}

\section{Introduction}
The study of random matrix theory (RMT) and orthogonal polynomials (OPs) for varying weights is often related to Painlev\'{e} equations. For example, the properties of Hankel determinants which are always playing the most fundamental role in RMT, as well as the asymptotic behavior of the recurrence coefficients with respect to OPs, can be found by the analysis of the associated particular Painlev\'{e} equations. Below we include a few related research works.
{\small
\begin{table}[H]
\centering
\begin{tabular}{c|c}
\hline
\hline
{\color{black}Equations}  &\parbox[t]{10cm}{~~~~~~~~~~~~~~~~~~~~~~~~ {\color{black}W}eight functions}\\
\hline
  Painlev\'{e} II & \parbox[t]{11cm}{${\rm e}^{\frac{z^3}{3}+tz},~z^3<0,~t\in\mathbb{R}$ [\cite{Mag1}].}\\
  Painlev\'{e} III & \parbox[t]{11cm}{$z^{\alpha}{\rm e}^{-z-t/z},~z,\alpha,t\in\mathbb{R}^{+}$ [\cite{CI1,MC1}].}\\
  Painlev\'{e} IV & {\color{black}\parbox[t]{11cm}{$z^{\lambda}{\rm e}^{-z^2+tz},~z\in\mathbb{R}^+,~\lambda>-1,~t\in\mathbb{R}$ [\cite{BvA,CJ1,FvAZ}];~${\rm e}^{-z^4+tz^2},~z,t\in\mathbb{R}$ [\cite{Mag1}]; $|z|^{2\lambda+1}{\rm e}^{-z^4+tz^2},~z,t\in\mathbb{R},~\lambda>-1$ [\cite{CJK1,FvAZ}]; $|z-t|^{\gamma}{\rm e}^{-z^2},~z,t\in\mathbb{R},~\gamma\geq0$ [\cite{CF1}].}}\\
  Painlev\'{e} V & {\color{black}\parbox[t]{11cm}{$(1-z)^{\alpha}(1+z)^{\beta}{\rm e}^{-tz},z\in[-1,1],t,\alpha,\beta\in\mathbb{R}^{+}$ [\cite{BC1,ZBCZ}]; $z^{\alpha}(1-z)^{\beta}{\rm e}^{-t/z},z\in[0,1],t,\alpha,\beta\in\mathbb{R}^{+}$[\cite{CD1}]; $z^{\alpha}(z+t)^{\beta}{\rm e}^{-z},z,t,\alpha,\beta\in\mathbb{R}^{+}$[\cite{FW1}]; $(1-z^2)^{\alpha}{\rm e}^{-t/z^2},z\in[-1,1],t,\alpha\in\mathbb{R}^{+}$[\cite{MC3}].}}\\
  Painlev\'{e} VI & {\color{black}\parbox[t]{11cm}{$z^{\alpha}(1-z)^{\beta}(t-z)^{\gamma},~z\in[0,1],~t,\alpha,\beta,\gamma\in\mathbb{R}^{+}$, [\cite{BCM,DZ1}].}}\\
\hline
\hline
\end{tabular}
\caption{Painlev\'{e} equations for varying weights}
\end{table}
}
In the present work, we mainly focused on the relationship between Painlev\'{e} equations and the singularly perturbed Jacobi weight,
\begin{equation}\label{weight}
w(z)=\left(1-z^2\right)^{\alpha}{\rm e}^{-\frac{t}{z^2-k^2}},~~~z,k\in[-1,1],~\alpha,t>0.
\end{equation}
The randomness of singularities $k\in[-1,1]$ is the motivation of this work.
\section{Preliminaries and the main result}
Let $P_n(z)$ be the monic orthogonal polynomials of degree $n$ in $z$ associated with the weight \eqref{weight}, i.e.
\begin{equation*}
\begin{aligned}
\int_{-1}^{1}P_{m}(z;t)P_{n}(z;t)w(z;t)dz=h_{n}(t)\delta_{m,n},\quad m,n=0,1,2,\ldots,
\end{aligned}
\end{equation*}
where
\begin{equation*}
\begin{aligned}
P_{n}(z;t)=z^{n}+\mathrm{p}(n,t)z^{n-2}+\ldots
\end{aligned}
\end{equation*}
In what follows, unless it is required, we will suppress the $t$ dependence for convenience. The monic orthogonal polynomials satisfy the three term recurrence relation
\begin{equation}\label{add-1}
\begin{aligned}
zP_n(z)=P_{n+1}(z)+\beta_nP_{n-1}(z),
\end{aligned}
\end{equation}
with initial conditions $P_{0}(x,t)=1,~ \beta_0P_{-1}(x,t)=0$. From the orthogonality, it is easy to find that
\begin{equation}\label{1.3}
\beta_{n}(t)=\mathrm{p}(n,t)-\mathrm{p}(n+1,t)=\frac{h_{n}(t)}{h_{n-1}(t)},
\end{equation}
and a telescopic sum gives
\begin{equation*}
\begin{aligned}
\sum\limits_{j=0}^{n-1}\beta_{j}(t)=-\mathrm{p}(n,t).
\end{aligned}
\end{equation*}
The lowering and raising operators for orthogonal polynomials, see e.g. [\cite{BC1,CD1,CI2,CI1,CZ1}]
\begin{equation*}
\begin{aligned}
&P'_{n}(z)=-B_{n}(z)P_{n}(z)+\beta_{n}A_{n}(z)P_{n-1}(z),\\
&P'_{n-1}(z)=\left[B_{n}(z)+v'(z)\right]P_{n-1}(z)-A_{n-1}(z)P_{n}(z).
\end{aligned}
\end{equation*}
where
\begin{subequations}
\begin{align}
&A_{n}(z):=\frac{1}{h_{n}}\int^{1}_{-1}\frac{v'(z)-v'(y)}{z-y}P^{2}_{n}(y)w(y)dy,\label{2.3}\\
&B_{n}(z):=\frac{1}{h_{n-1}}\int^{1}_{-1}\frac{v'(z)-v'(y)}{z-y}P_{n}(y)P_{n-1}(y)w(y)dy. \label{2.4}
\end{align}
\end{subequations}
and $v(z):=-\ln w(z)$. Note that the functions $A_{n}(z)$ and $B_{n}(z)$ are not independent but satisfy the following supplementary conditions for $z\in \mathbb{C}\cup \{\infty\}$
\begin{align}
B_{n+1}(z)+B_{n}(z)=zA_{n}(z)-v'(z)~~~~~~~~~~~~~~~~~~~~~~\tag{$S_1$}\\
1+z(B_{n+1}(z)-B_{n}(z))=\beta_{n+1}(t)A_{n+1}(z)-\beta_{n}(t)A_{n-1}(z)\tag{$S_2$}\\
B^{2}_{n}(z)+v'(z)B_{n}(z)+\sum ^{n-1}_{j=0}A_{j}(z)=\beta_{n}(t)A_{n}(z)A_{n-1}(z)~~~~~~\tag{$S_2'$}
\end{align}
The main result of this work is as follows.
\begin{thm}
Defining $R_{n}(t):=\frac{2t}{h_{n}(t)}\int^{1}_{-1}\frac{1}{y^{2}-k^{2}}P^{2}_{n}(y)w(y)dy$ and {\color{black}letting
$$\Phi_n(t):=\frac{R_{n}(t)+2n+2\alpha+1}{2n+2\alpha+1},$$}
then $\Phi_n(t)$ satisfies the following Painlev\'{e} V equation [\cite{T-1}].
\begin{equation*}
\begin{aligned}
\Phi_n''(t)=&\frac{(3\Phi_n(t)-1){\Phi_n'(t)}^2}{2\Phi_n(t)(\Phi_n(t)-1)}-\frac{\Phi_n'(t)}{t}+\frac{(\Phi_n(t)-1)^2}{t^2}\left(\gamma\Phi_n(t)+\frac{\delta}{\Phi_n(t)}\right)\\
&+\frac{\varepsilon\Phi_n(t)}{t}+\frac{\eta\Phi_n(t)(\Phi_n(t)+1)}{\Phi_n(t)-1}\\
\end{aligned}
\end{equation*}
where
\begin{equation*}
\gamma=\frac{(2n+2\alpha+1)^2}{8},~~~\delta=-\frac{1}{8},~~~\varepsilon=-\frac{\alpha}{k^2},~~~\eta=-\frac{1}{2k^4}.
\end{equation*}
\end{thm}

\section{Proof of the main result}
\noindent According to definitions of $A_{n}(z)$ and $B_n(z)$ in \eqref{2.3} and \eqref{2.4}, with the aid of the orthogonality relation \eqref{add-1} and some integration techniques, we have
\begin{subequations}
\begin{align}
&A_{n}(z)=\frac{a_{n}(t)}{1-z^{2}}+\frac{a_{n}(t)-2n-2\alpha-1}{z^{2}-k^{2}}+\frac{k^{2}R_{n}(t)}{(z^{2}-k^{2})^{2}},\label{A}\\
&B_{n}(z)=\frac{zb_{n}(t)}{1-z^{2}}+\frac{z(b_{n}(t)-n)}{z^{2}-k^{2}}+\frac{zr_{n}(t)}{(z^{2}-k^{2})^{2}},\label{B}
\end{align}
\end{subequations}
where
\begin{equation*}
\begin{aligned}
&R_{n}(t):=\frac{2t}{h_{n}}\int^{1}_{-1}\frac{1}{y^{2}-k^{2}}P^{2}_{n}(y)w(y)dy,\\
&a_{n}(t):=\frac{2\alpha}{h_{n}}\int^{1}_{-1}\frac{1}{1-y^{2}}P^{2}_{n}(y)w(y)dy,\\
&r_{n}(t):=\frac{2t}{h_{n-1}}\int^{1}_{-1}\frac{y}{y^{2}-k^{2}}P_{n}(y)P_{n-1}(y)w(y)dy,\\
&b_{n}(t):=\frac{2\alpha}{h_{n-1}}\int^{1}_{-1}\frac{y}{1-y^{2}}P_{n}(y)P_{n-1}(y)w(y)dy.
\end{aligned}
\end{equation*}

Substituting \eqref{A} and \eqref{B} in $S_1$, then comparing the coefficients at $\mathcal{O}(\frac{z}{1-z^2})$, {\color{black}$\mathcal{O}(\frac{z}{(z^2-k^2)^2})$,} we get
\begin{equation*}
\begin{aligned}
&b _{n+1}+b _{n}=a_{n}-2\alpha,\\
&r_{n+1}+r_{n}=k^{2}R_{n}+2t.
\end{aligned}
\end{equation*}
Similarly, substituting \eqref{A} and \eqref{B} into $S_2$, we have
\begin{subequations}
\begin{align}
&b _{n+1}-b _{n}=\beta_{n+1}a_{n+1}-\beta_{n}a_{n-1},\label{s2-1}\\
&r_{n+1}-r_{n}=\beta_{n+1}R_{n+1}-\beta_{n}R_{n-1}, \label{s2-2}\\
&r_{n+1}-r_{n}+(k^{2}-1)(b_{n+1}-b_{n})-k^{2}=\beta_{n}(2n+2\alpha-1)-\beta_{n+1}(2n+2\alpha+3).\label{s2-3}
\end{align}
\end{subequations}
We rewrite \eqref{s2-3} as
\begin{equation*}
\begin{aligned}
(k^{2}-1)(b_{n+1}-b_{n})=\beta_{n}(2n+2\alpha-1)-\beta_{n+1}(2n+2\alpha+3)-(r_{n+1}-r_n)+k^2,
\end{aligned}
\end{equation*}
then we substituting \eqref{s2-2} into the above equation, we have
\begin{equation*}
\begin{aligned}
(k^{2}-1)(b_{n+1}-b_{n})=\beta_{n}(R_{n-1}+2n+2\alpha-1)-\beta_{n+1}(R_{n+1}+2n+2\alpha+3)+k^2.
\end{aligned}
\end{equation*}
Letting $k\rightarrow0$, we can see that
\begin{equation}\label{yanjin-3}
\begin{aligned}
b_{n+1}-b_{n}=\beta_{n}(R_{n-1}+2n+2\alpha-1)-\beta_{n+1}(R_{n+1}+2n+2\alpha+3),
\end{aligned}
\end{equation}
comparing \eqref{s2-1} with \eqref{yanjin-3}, we obtain
\begin{equation}\label{yanjin-4}
\begin{aligned}
a_n=R_{n}+2n+2\alpha+1.
\end{aligned}
\end{equation}
Substituting (\ref{s2-1}) and (\ref{s2-2}) into (\ref{s2-3}), we find
\begin{equation*}
\begin{aligned}
r_{n+1}-r_{n}+(k^{2}-1)(b_{n+1}-b_{n})-k^{2}&=\beta_{n}(2n+2\alpha-1)-\beta_{n+1}(2n+2\alpha+3)\\
&=\beta_{n}(2n+2\alpha-1)-\beta_{n+1}(2n+2\alpha+1)-2\beta_{n+1}
\end{aligned}
\end{equation*}
A telescopic sum gives
\begin{equation}\label{zhu2.23}
\begin{aligned}
r_{n}+(k^{2}-1)b_{n}-nk^{2}=-\beta_{n}(2n+2\alpha+1)+2\mathrm{p}(n,t).
\end{aligned}
\end{equation}

Substituting \eqref{A} and \eqref{B} into $S_{2}^{'}$, $z^2$ can be replaced with $(z^2-k^2)+k^2$ or $(z^2-1)+1$, then comparing the coefficients at $\mathcal{O}(\frac{z}{1-z^2})$, $\mathcal{O}(\frac{z}{z^2-k^2})$, $\mathcal{O}(\frac{z}{(1-z^2)^2})$, $\mathcal{O}(\frac{z}{(z^2-k^2)^2})$,
$\mathcal{O}(\frac{z}{(1-z^2)(z^2-k^2)})$, $\mathcal{O}(\frac{z}{(1-z^2)(z^2-k^2)^2})$, $\mathcal{O}(\frac{z}{(z^2-k^2)^3})$, $\mathcal{O}(\frac{z}{(z^2-k^2)^4})$, we obtain
\begin{numcases}{}
b_{n}^{2}+2\alpha b_{n}=\beta_{n}a_{n}a_{n-1}, \label{s3-1}\\
r_{n}^{2}-2tr_{n}=k^{2}\beta_{n}R_{n}R_{n-1},\label{s3-2}\\
b_{n}^{2}-2n(b_n+\alpha)+\sum\limits_{j=0}^{n-1}a_{j}=0,\label{s3-3}\\
2b_{n}(r_{n}-t)+2\alpha r_{n}=\beta_{n}(a_{n}R_{n-1}+a_{n-1}R_{n}),\label{s3-4}\\
k^{2}(b_{n}-n)^{2}-2t(b_{n}-n)+2r_{n}(b_{n}-n)+k^{2}\sum\limits_{j=0}^{n-1}R_{j}=\beta_{n}R_nR_{n-1},\label{s3-5}\\
2k^{2}(b_{n}-n)(b_n+\alpha)+2b_{n}(r_{n}-t)+2\alpha r_{n}=\beta_{n}(a_{n-1}R_n+a_{n}R_{n-1}),\label{s3-6}\\
r_n^2-2tr_n+2k^2(b_n-n)(r_n-t)=2k^2\beta_{n}R_nR_{n-1}.\label{s3-7}
\end{numcases}
or
\begin{numcases}{}
b_n^2+2\alpha b_n=\beta_na_na_{n-1},\label{s3'-1}\\
r_n^2-2tr_n=k^2\beta_nR_nR_{n-1},\label{s3'-2}\\
b_n^2+2\alpha b_n=\sum_{j=0}^{n-1}a_j,\label{s3'-3}\\
2b_nr_n+2\alpha r_n-2\alpha tb_n=k^2\beta_n(a_nR_{n-1}+a_{n-1}R_n), \label{s3'-4}\\
k^2(b_n-n)^2-2t(b_n-n)-2r_n(n+\alpha)+2\alpha tb_n+k^2\sum_{j=0}^{n-1}R_j=\beta_{n}R_nR_{n-1},\label{s3'-5}\\
2(b_n+\alpha)(b_n-n)=\beta_n(a_{n-1}R_n+a_nR_{n-1}),\label{s3'-6}\\
r_n^2-2tr_n+2k^2(b_n-n)(r_n-t)=2k^2\beta_{n}R_nR_{n-1}.\label{s3'-7}
\end{numcases}
Substituting  \eqref{s3-4} into \eqref{s3-7}
\begin{equation}\label{zhu2.32}
\begin{aligned}
r_{n}^{2}-2k^{2}(n+\alpha)r_{n}+2k^{2}tn-2tr_{n}+\beta_{n}k^{2}R_{n}(2n+2\alpha-1)\\
+\beta_{n}k^{2}R_{n-1}(2n+2\alpha+1)=0,
\end{aligned}
\end{equation}
then substituting \eqref{s3-2} into \eqref{zhu2.32}, we can obtain the expression of $\beta_{n}$
\begin{equation}\label{2.33}
\begin{aligned}
\beta_{n}=\frac{2k^{2}(n+\alpha)r_{n}-2k^{2}tn+2tr_{n}-r_{n}^{2}}{k^{2}R_{n}(2n+2\alpha-1)}-
\frac{(2n+2\alpha+1)(r_{n}^{2}-2tr_{n})}{k^{2}R_{n}^{2}(2n+2\alpha-1)}.
\end{aligned}
\end{equation}

Taking the derivative of $h_n(t)$ with respect to $t$,
\begin{align*}
-\int_{-1}^{1}\frac{1}{z^{2}-k^{2}}P_{n}^{2}(z;t)w(z;t)dx=h'_{n}(t),
\end{align*}
it will be easy to see that
\begin{equation*}
\begin{aligned}
2t\frac{d}{dt}\mathrm{ln}h_{n}(t)=-R_{n}(t),
\end{aligned}
\end{equation*}
because of the definition of $R_{n}(t)$. Using \eqref{1.3}, we have
\begin{align*}
2t\frac{d}{dt}\mathrm{ln}\beta_{n}(t)=R_{n-1}(t)-R_{n}(t),
\end{align*}
hence
\begin{equation}\label{3.2}
\begin{aligned}
2t\beta'_{n}(t)=\beta_{n}(t)R_{n-1}(t)-\beta_{n}(t)R_{n}(t).
\end{aligned}
\end{equation}
Based on the orthogonality, $\int_{-1}^{1}P_{n}(z)P_{n-2}(z)w(z)dz=0$, taking a derivative with respect to $t$, we have
\begin{equation}\label{3.3}
\begin{aligned}
\frac{d}{dt}\mathrm{p}(n,t)=\frac{1}{h_{n-2}(t)}\int_{-1}^{1}\frac{1}{z^{2}-k^{2}}P_{n}(z;t)P_{n-2}(z;t)w(z;t)dz.
\end{aligned}
\end{equation}
Using \eqref{add-1} and \eqref{1.3}, i.e.
\begin{equation}\label{3.4}
\begin{aligned}
\frac{P_{n-2}(z;t)}{h_{n-2}(t)}=\frac{zP_{n-1}(z;t)}{h_{n-1}(t)}-\frac{P_{n}(z;t)}{h_{n-1}(t)},
\end{aligned}
\end{equation}
substituting \eqref{3.4} into \eqref{3.3}, and bear in mind $\beta_n=h_n/h_{n-1}$, we have
\begin{equation}\label{3.5}
\begin{aligned}
2t\frac{d}{dt}\mathrm{p}(n,t)=r_{n}-\beta_{n}R_{n}(t).
\end{aligned}
\end{equation}

Taking a derivative in \eqref{zhu2.23} and substituting \eqref{3.2} and \eqref{3.5} into it,
\begin{equation*}
\begin{aligned}
2tr'_{n}+2(k^{2}-1)tb'_{n}=2r_{n}+(2n+2\alpha-1)\beta_{n}R_{n}-(2n+2\alpha+1)\beta_{n}R_{n-1},
\end{aligned}
\end{equation*}
substituting \eqref{s3-2} into above equation,
\begin{equation*}
\begin{aligned}
2tr'_{n}+2(k^{2}-1)tb'_{n}=2r_{n}+(2n+2\alpha-1)\beta_{n}R_{n}-
\frac{(2n+2\alpha+1)(r_{n}^{2}-2tr_{n})}{k^{2}R_{n}},
\end{aligned}
\end{equation*}
comparing the equations \eqref{s3-3} and \eqref{s3'-3}, with the aid of \eqref{2.33}, we have
\begin{equation}\label{p-1.1}
\begin{aligned}
2k^2tr'_{n}(t)=&2\left[k^2(n+\alpha+1)+t\right]r_n(t)-\frac{2(2n+2\alpha+1)\left(r_{n}^{2}(t)-2tr_{n}(t)\right)}{R_n(t)}\\
&-r_n^2(t)-2k^2nt.
\end{aligned}
\end{equation}
Next, combining \eqref{3.2} and \eqref{s3-2}, we get
\begin{equation*}
\begin{aligned}
2t\beta'_{n}=\frac{r_{n}^{2}-2tr_{n}}{k^{2}R_{n}}-\beta_{n}R_{n},
\end{aligned}
\end{equation*}
Substituting \eqref{2.33} into above equation, and using \eqref{p-1.1} to eliminate the terms involving $r'_{n}(t)$, we have
{\small\begin{equation*}
\begin{aligned}
&0=\big[2tR_n+2k^2(n+\alpha+1)R_n+k^2R_{n}^2-2r_n(2n+2\alpha+1+R_n)+2t(2n+2\alpha+1)\\
&-2k^2tR_{n}'\big]\big[2(2n+2\alpha+1)(2t-r_n)r_n+\big(2tr_n+2k^2nr_n+2k^2\alpha r_n-r_n^2-2k^2nt\big)R_n\big]
\end{aligned}
\end{equation*}}
\noindent It yields two equations, however the algebraic equation for $R_n(t)$ and $r_{n}(t)$ does not hold. This can be checked by taking special values. So we obtain
\begin{equation}\label{p-1.2}
\begin{aligned}
2k^2tR'_{n}(t)=&2\left[k^2(n+\alpha+1)+t\right]R_n(t)-2r_n(t)(2n+2\alpha+1+R_n(t))\\
&+k^2R_n^2(t)+2(2n+2\alpha+1)t.
\end{aligned}
\end{equation}
Note that equations \eqref{p-1.1} and \eqref{p-1.2} are also called the coupled Riccati equations. Solving for $r_n(t)$ from \eqref{p-1.2} and substituting it into \eqref{p-1.1}, we obtain
\begin{equation*}
\begin{aligned}
&8k^4t^2R_n(2n+2\alpha+1+R_n)R_n''-4k^4t^2(4n+4\alpha+2+3R_n)R_n'^2\\
&+8k^4tR_n(2n+2\alpha+1+R_n)R_n'-k^4R_n^5-2k^4(2n+2\alpha+1)R_n^4\\
&-4\left[k^4(n+\alpha)(n+\alpha+1)-t^2-2k^2\alpha t\right]R_n^3+16t(2n+2\alpha+1)(t+k^2\alpha)R_n^2\\
&+4t(2n+2\alpha+1)^2(5t+2k^2\alpha)R_n+8t^2(2n+2\alpha+1)^3=0.
\end{aligned}
\end{equation*}
Applying the linear transformation, the result will be yielded immediately.

\section{Acknowledgements}
C. Li acknowledges the support  of the National Natural Science Foundation of China under Grant no. 11571192 and K. C. Wong Magna Fund in Ningbo University.

M. Zhu and Y. Chen would like to thank the Science and Technology Development Fund of the Macau SAR for generous support in providing FDCT 023/2017/A1. They would also like to thank the University of Macau for generous support via MYRG 2018-00125 FST.

\end{document}